\documentclass[aps,pra, twocolumn]{revtex4}
\usepackage{graphicx}
\usepackage{bm}
\usepackage{amsmath}
\usepackage[colorlinks=true,allcolors=blue]{hyperref}

\providecommand{\bra}[1]{\langle #1 \rvert}
\providecommand{\ket}[1]{\lvert #1 \rangle}
\providecommand{\braket}[2]{\langle #1 \rvert #2 \rangle}

\providecommand{\be}{\begin{equation}}
\providecommand{\ee}{\end{equation}}
\providecommand{\ba}{\begin{eqnarray}}
\providecommand{\ea}{\end{eqnarray}}

\usepackage{mathbbol}
\usepackage{amsmath}
\usepackage{amssymb}
\usepackage{amsthm}
\newtheorem{theorem}{Theorem}

\newcommand{\ie}{\textit{i.e.} }

\begin{document}

\title{Quantifying quantum invasiveness}

\author{ Saulo V. Moreira$^{1,2}$ and Marcelo {Terra Cunha}$^2$}

\affiliation{$^1$Centro de Ci\^encias Naturais e Humanas, Universidade Federal do ABC - UFABC, Santo Andr\'e, Brazil}
\affiliation{$^2$ Departamento de Matem\'atica Aplicada, Instituto de Matem\'atica, Estat\'istica e Computa\c{c}\~ao Cient\'ifica, Universidade Estadual de Campinas - Unicamp, 13084-970, Campinas, Brazil}

\begin{abstract}
We propose a resource theory of the quantum invasiveness of general quantum operations, i.e., those defined by quantum channels in Leggett-Garg scenarios.
We are then able to compare the resource-theoretic framework of quantum invasiveness to the resource theory of coherence.
We also show that the Fisher information is a quantifier of quantum invasiveness.
This result allows us to establish a direct connection between the concept of quantum invasiveness and quantum metrology, by exploring the utility of the definition of quantum invasiveness in the context of metrological protocols. 

\end{abstract}
\vskip2pc

\maketitle

\section{Introduction} \label{introduction}

Recently, substantial efforts have been undertaken by the community to elucidate and quantify nonclassical features of quantum theory by using the mathematical framework of \textit{resource theories} \citep{Coecke}. Perhaps the most studied and developed example of a nonclassical property in the context of resource theories is the quantification of entanglement \citep{Horodecki}. Similarly, the resource theory of quantum coherence \citep{Baumgratz} has been proposed and widely explored in the last few years \citep{Streltstov}, providing new insights and leading to applications in a variety of topics on quantum information and technologies, such as quantum metrology \citep{Marvian}, quantum thermodynamics \citep{Misra} and quantum biology \citep{Li}. In this spirit, the proposition of resource theories of nonclassical aspects would potentially permit us to improve our understanding of these aspects, helping to establish equivalences and connections with other nonclassical features of quantum theory. Naturally, new applications and potentialities of nonclassicality in resource-theoretic frameworks could be explored as well.

Some nonclassical aspects of quantum theory are usually defined as those not satisfying a worldview referred to as \textit{macrorealism}. 
This notion was introduced by Leggett and Garg in the 1980s and was associated with the intuitive fact that \textit{macroscopic} objects - in this context understood as those governed by classical laws - are consistent with definite values for their properties at all instants of time, and that measurements performed on these objects cannot have an effect on these values \citep{LG}. Aiming to propose a test capable of ruling out macrorealism in physical systems, the authors proposed the so-called \textit{Leggett-Garg inequality} (LGI), the derivation of which was based on the two aforementioned assumptions. The violation of the inequality would then permit us to rule out macrorealism. However, the precise meaning of the violation of the LGI with regards to these assumptions has been the subject of debate since the inequality has been proposed, regaining relevance in the last few years \citep{Maroney, Emary, Clemente, Kumari, Moreira, Clemente2}. 

Amidst this debate, there have been propositions of alternative conditions for macrorealism, such as the \textit{no-signaling in time condition} (NSIT) \citep{Kofler2}. 
This condition puts forward the macrorealist assumption related to the null effect that the measurement process is expected to have on a physical system, as it compares the statistics of the measurement of a chosen observable in two experiments: one in which a non-selective previous measurement is performed at time $ t_1 $  before a later one at $ t_2 > t_1 $, and another where the measurement at $ t_1 $ is absent (non-selective, here, means that irrespectively of the result of the measurement, the physical system will continue its history). 
In a macrorealistic world, one expects the statistics of the measurement at $ t_2 $ to be the same in both experiments, since measurements can be carried out with arbitrarily small disturbance to the system's state. Later, by considering a scheme where three measurements can be performed, it was shown in Ref. \citep{Clemente} that the fulfilment of sets of specific NSIT conditions may be \textit{necessary and sufficient} for macrorealism. This was shown by considering  an \textit{underlying probability} associated with a scenario where three measurements are performed from which all the probabilities (associated, for instance, with experiments where only one or two out of the three predefined measurements are carried out) can be obtained by marginalizing the underlying probability \cite{Fine, Amaral, Mansfield}. Similar conditions were used in the context of the LGI by Maroney and Timpson \citep{Maroney} to show that LGI violation can always be related to a notion of the measurement disturbing the system's evolution in a nonclassical fashion. We will refer to this nonclassical concept, from now on, as \textit{quantum invasiveness}, or simply \textit{invasiveness}.

Recent works have explored connections of the violation of NSIT conditions \citep{Frowis} and LGIs \citep{Moreira2} with \textit{quantum metrology} \citep{Giovannetti, Paris,Escher, Escher1}, as well as suggested a unified approach to contextuality and violations of macrorealism \citep{Mansfield}.

In this paper, we seek to propose a resource theory of operations, the \textit{resource theory of quantum invasiveness}. 
Also, within the resource-theoretic framework, we are able to explore the connections as well as the contrasts between the nonclassical concepts of invasiveness and coherence.
This paper is organized as follows. 
We discuss briefly the resource theoretical framework in section \ref{rt}, as well as the resource theory of coherence in section \ref{coherence}.
In section \ref{qinvasiveness}, we present the scenario as well as the definition of invasiveness, and the resource theory of quantum invasiveness is introduced section \ref{rtinvasiveness}. 
Finally, in section \ref{QuantInvasiveness} we outline the definition of the Fisher information and quantum Fisher information  \cite{Fisher, Cramer, Rao} in the context of protocols for parameter estimation, as well as show that the Fisher information is an invasiveness quantifier. This result is discussed in section \ref{discussion}, as well as the relationship between invasiveness and coherence with the aid of the resource-theoretic framework.

\section{The resource-theoretic framework} \label{rt}
 
As mentioned above, one of the most remarkable uses of resource theories is related to the quantification of a given \textit{resource}, which can be useful in performing certain tasks. As pointed out in Ref. \citep{Coecke}, a way to define resource theories is by sorting a given a set of experimental interventions (preparations, transformations and measurements, for instance) into the \textit{free} and the \textit{costly} interventions. Presumably, one should be able to use the elements of the free set unlimitedly and in any combination. The costly elements, on the other hand, are the resources. In this way, it is expected that a resource theory describes the structure induced on the resources, given access to the free set. Resource theories can be formalized by defining them as \textit{symmetric monoidal categories} \citep{Coecke2}. Hence, the \textit{objects} of the category can be composed both in parallel and in sequence \citep{Coecke}. 

Here, we only provide a basic description of the requirements to define a resource theory. As mentioned above, the elements of the free set must be specified in order to define it. The objects of a resource theory possessing no value are called \textit{free resources}. In turn, \textit{free transformations} are the transformations between two objects which can be implemented without any cost. Thus, it is expected that free transformations map free objects into free objects. Logically, free resources are therefore expected to remain costless after being subjected to a free transformation. 

\subsubsection{Example: Resource theory of entanglement}

In a standard approach to the resource theory of entanglement, free transformations are defined as local operations and classical communication (LOCC), whilst free states are considered to be separable states \citep{Horodecki}. 
Consistently, entangled states cannot be generated by LOCC. 
Resource theories of this sort, where \textit{states} can be identified as the relevant resource, are commonly referred to as \textit{resource theories of states}.
An example of \textit{quantifier} of the resource entanglement of a state $\rho$ is the \textit{relative entropy of entanglement}, defined as \citep{Vidal,Vedral,Schumacher,Horodecki}
\begin{equation}\label{entropy}
E_R = \inf_{\sigma \in \Omega} {\rm Tr} \rho(\log\rho - \log\sigma),
\end{equation}
where $\Omega$ denotes the set of separable states.

\section{Coherence}\label{coherence}

Before moving on, it will be useful to briefly review the resource theory of coherence as proposed in Ref. \citep{Baumgratz}, since it can be connected to the resource theory of invasiveness, as we will discuss later.

Given a basis $\{\ket{i}\}$, the incoherent states (free objects) $\mathcal{I}$ are defined as the diagonal states in this basis. 
Free operations, in turn, are those leading a the set of diagonal state into itself.
Therefore, they cannot generate coherence.

An example of a coherence quantifier satisfying the conditions above is the $l_1$-\textit{norm of coherence} \citep{Baumgratz}, 
\begin{equation}\label{LNorm}
D_{l_1} = \sum_{ij,j\neq i} |\rho_{ij}|,
\end{equation}
where $\rho_{ij}$ are the off-diagonal elements of a given state $\rho$.

In the following, we will present the definition of the nonclassical concept of quantum invasiveness of a quantum channel $\Phi$ in Leggett-Garg scenarios.
Based on this, we will propose a resource theory of quantum invasiveness, as well as identify how it can be related to the resource theory of coherence.

\section{Quantum invasiveness} \label{qinvasiveness}

\begin{figure}[h]
\centering 
\includegraphics[width=8.5cm]{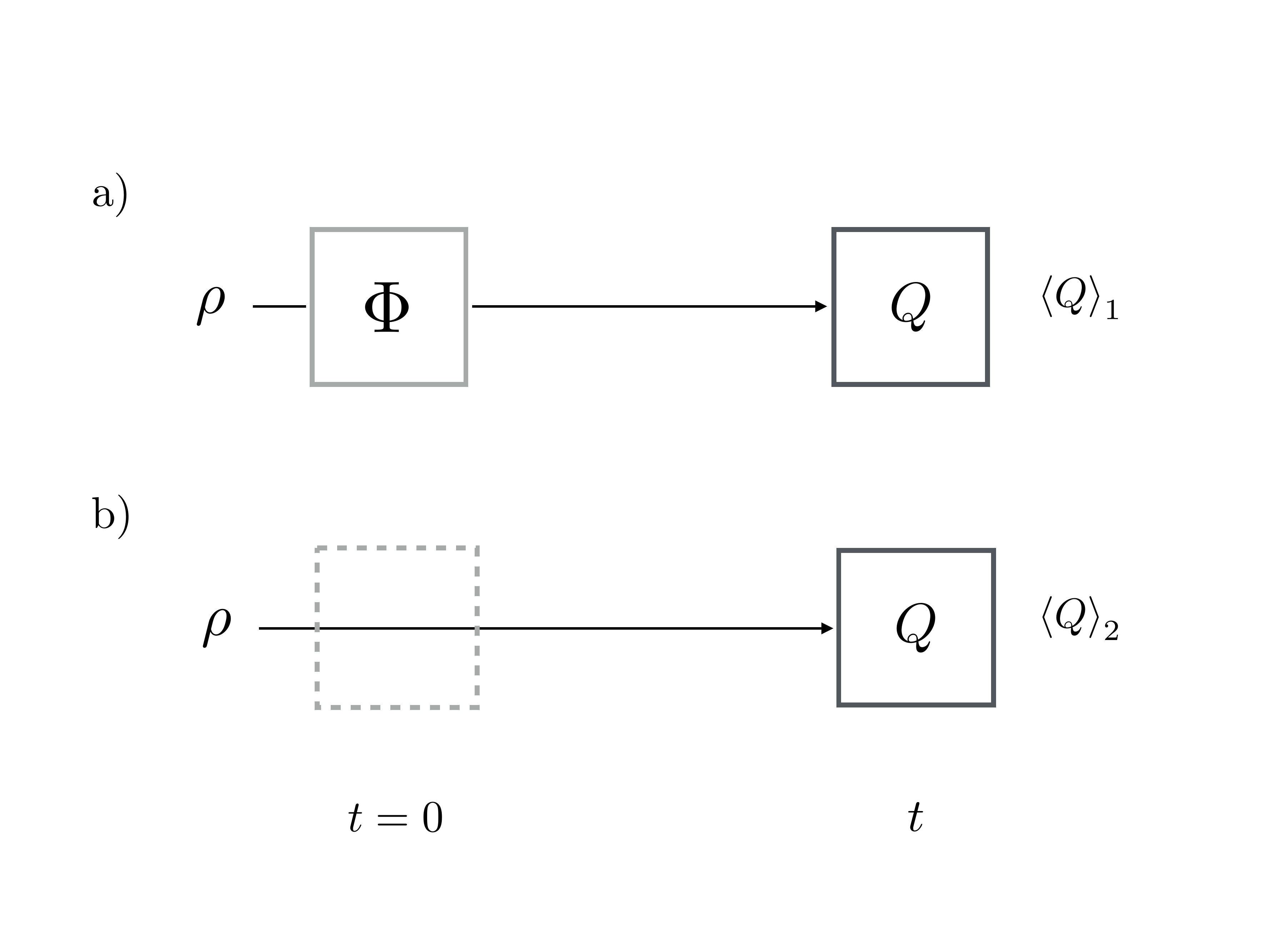}
\caption{Schematic representation of the scenario - a) Experiment 1: the state $\rho$ is subjected to the operation $\Phi$ at $t=0$ and the observable $Q$ is measured at $t$. By repeating this experiment many times, the expected value $ \left<Q\right>_1 $ is obtained. b) Experiment 2: the empty box at $t=0$ represents the fact that the operation $\Phi$ is not applied. At $t$, $Q$ is measured. As before, many realizations of the experiment allows the obtention of $ \left<Q\right>_2 $.  }\label{Scenario}
\end{figure}

A generalization of the concept of invasiveness has been recently proposed in Ref. \citep{NewKnee}. 
The nonclassical notion of measurement invasiveness was extended to \textit{invasiveness of a quantum operation}, represented by a quantum channel. 
In order to introduce this concept, we first describe the associated scenario, schematically shown in Fig. \ref{Scenario}. 
We consider a state $\rho$, a CPTP quantum channel $\Phi$ associated with the Kraus operators $ \{ K_l \} $ satisfying~$\sum_l K^\dagger_lK_l = \mathbb{1}$, and an observable $Q$. 
Trace preserving condition is consistent with the non-selectiveness imposed on measurements. 
Such condition can be drop, but we prefer to impose it and keep things simpler.

In an experiment labeled as Experiment 1, $\Phi$ is applied to the state $\rho$ at $t=0$: $ \rho \mapsto  \sum_l K_l \rho K^\dagger_l  $. 
Then, the observable $Q$ is measured at $t$. 
This experiment is repeated many times, in such way that the expected value of the observable $Q$, $ \left<Q \right>_1 $, is obtained. 
In a second experiment, the Experiment 2, $\rho$ is not subjected to $\Phi$, and the observable $Q$ is measured at $t$. 
The experiment is repeated several times as well, so that the expected value of $Q$, $\left<Q\right>_2$, can be evaluated.
\\

Considering the scenario described above, $\rho$ and $Q$ generate a witness to the invasiveness of the quantum operation $\Phi$, $ W $,  defined as~\cite{Schild, Knee, NewKnee}:
\begin{equation}\label{Winvasiveness}
W \equiv \left<Q\right>_1 - \left<Q\right>_2 = {\rm Tr}(Q\Phi(\rho) - Q\rho ).
\end{equation}
One may question the fact that $W$ may not be zero as a result of classical disturbances or errors present in the experiments. 
To rule out this possibility, a strategy consisting of a protocol with control experiments was proposed in Refs. \citep{NewKnee, Knee}. 
These control experiments allow determining $W$ by performing the experiments described above for the eigenstates of the observable $Q$ as inputs. 
Eigenstates are associated with definite values of the quantity represented by the observable $Q$, and therefore, by performing  the control experiments, one can determine the classical disturbance.

In other words, the problem posed by the presence of classical errors can be experimentally tackled via control experiments \citep{Knee, NewKnee}.  Specifically, a control experiment has to be performed for each \textit{classical state}, defined as the eigenstates $\ket{q}$ of the observable $Q$. Thus, a control experiment corresponds to the scheme shown in Fig. \ref{Scenario} with the input state $\rho$ corresponding to a particular eigenstate $\ket{q}$, such that the corresponding value of the witness defined in Eq. \eqref{invasiveness} can be determined. By calling the values of the witness for each eigenstate $\ket{q}$ as input by $W_q$, one gets the condition $\min\{W_q\}\le W\le\max\{W_q\}$, instead of $W=0$, as the non-disturbance condition \citep{Knee}. Hence, $W \neq 0$ is a witness of invasiveness only if all $W_q = 0$. As an example, consider $Q=\sigma_z$ and $\Phi=\eta_z$, $\eta_z\rho = \sigma_z\rho \sigma_z$.
Before the measurement of $Q$ in Fig. \ref{Scenario}, we will consider a transformation $\eta_H$.
Therefore, in Experiment 1 sketched in Fig.  \ref{Scenario}, its combination with $\Phi$ will read $\eta_H\circ\eta_z\rho=H[\sigma_z\rho \sigma_z ]H$ with \citep{NewKnee}
\begin{equation}\label{unitary1}
H= \frac{1}{\sqrt{2}} \begin{pmatrix}
1 & 1 \\
1 & -1
\end{pmatrix}.
\end{equation}
Note that the transformation $\eta_H$ is always present, both in Experiment 1 of Fig. \ref{Scenario}, where $\Phi$ is also present, as well as in Experiment 2, where $\Phi$ is absent. 
The eigenstates of $Q$, defined as classical states, are $\{\ket{0},\ket{1}\}$. In this particular example, where $[Q,\sigma_z]=0$, we have that $W_0=W_1=0$. In turn, by taking $\rho=\ket{\vartheta}\bra{\vartheta}$ with $\ket{\vartheta}=\frac{1}{\sqrt{2}}(\ket{0}-\ket{1})$, we obtain $W=2$. We have therefore a simple example where all $W_i=0$ and invasiveness can be directly witnessed.


From Eq. \eqref{Winvasiveness} we are able to gain some intuition related to the fact that quantities depending on some notion of distance between $ \rho $ and $ \Phi(\rho)$ may be good candidates as quantifiers of invasiveness (see also \citep{Bilobran}). Based on the concepts and witnesses of nonclassicality discussed above, we present the definition of quantum invasiveness that we will consider from now on. 
\\
\\
{\bf Definition}. (Quantum invasiveness) \textit{A general quantum operation (represented by a quantum channel) is considered to be \textit{invasive} whenever it disturbs the physical system in a nonclassical way.}
\\
\\

In order for the definition above to be precise, it is necessary to specify the classical states and classical operations.
Indeed, since our main concern is the nonclassical effect of the invasiveness of operations, one may question whether a generalization of the scheme presented above, which would allow a classical operation instead of the absence of the operation - represented by the dashed box in Fig. \ref{Scenario}(b) - would be possible.
As we will see, the resource theory of quantum invasiveness introduced in the next section naturally allows us to include the possibility of the presence of classical operations in the dashed box, in contrast with the scheme above, which ultimately relies on control experiments to get rid of the classical disturbance.

\section{Quantifying invasiveness}\label{rtinvasiveness}

We now formulate a resource theory of operations for the invasiveness of transformations $\Phi$, given a measurement observable $Q$:
 \begin{itemize}
\item  \textit{{\bf Free states}}: The eigenstates $\ket{q}$ of the measurement observable $Q$, as well as their convex combinations, $\rho^C=\sum_k p_k \ket{q}\bra{q}$,  $\sum_k p_k =1$.  The set of free states will be denoted by $\Gamma$, i.e. $\rho^C \in \Gamma$.

\item \textit{{\bf Free operations}}: CPTP Quantum channels $\Phi_{\text{Free}}$ which can be expressed in Kraus representation using Kraus operators of the form $ K_l = \sum_i c_l(i)\ket{q_{j(i)}}\bra{q_i} $, where $j(i)$ is a function from the index set of the basis of $Q$, and $c_l(i)$ are coefficients \citep{Winter}.
\end{itemize}

The free-state set is the convex set generated by the eigenstates of $Q$, since, with respect to $Q$ measurements, they can be given a classical ontological interpretation, while their convex combinations receive a probabilistic interpretation based on ignorance.
For this reason, we also refer to any such $\rho^C$ as a classical state.
This definition of free states is in line with the notion of  \textit{eigenstate mixture macrorealism} in Ref. \citep{Maroney}.

In turn, free operations for a resource theory of quantum invasiveness must necessarily map a classical state into another classical state. 
The free operation defined above, which is an \textit{incoherent completely positive trace preserving map} \citep{Baumgratz}, satisfies this requirement. 
Specifically, in a Kraus representation, these operations are such that $ \rho_f =  \sum_l K_l \rho^C K^\dagger_l  \in \Gamma$   and all $\rho^C \in \Gamma$ \citep{Winter}. 
This restriction is such that even in the generalized version when one has access to individual measurement outcomes $\{K_l\}$ and non trace-preserving maps have to be used: $\rho^C \mapsto \frac{1}{{\rm{Tr}} \left(K_l \rho^C K^\dagger_l\right)}K_l \rho^C K^\dagger_l$, it is impossible to generate invasiveness from the free states.

This is a minimal set of elements allowing us to define a resource theory of invasiveness. 
Naturally, alternative resource theories of invasiveness can be defined by considering different free operation sets, in the same way that a resource theory for entanglement can also consider PPT-transformations instead of LOCC maps.

\begin{figure}[h]
\centering 
\includegraphics[width=9.2cm]{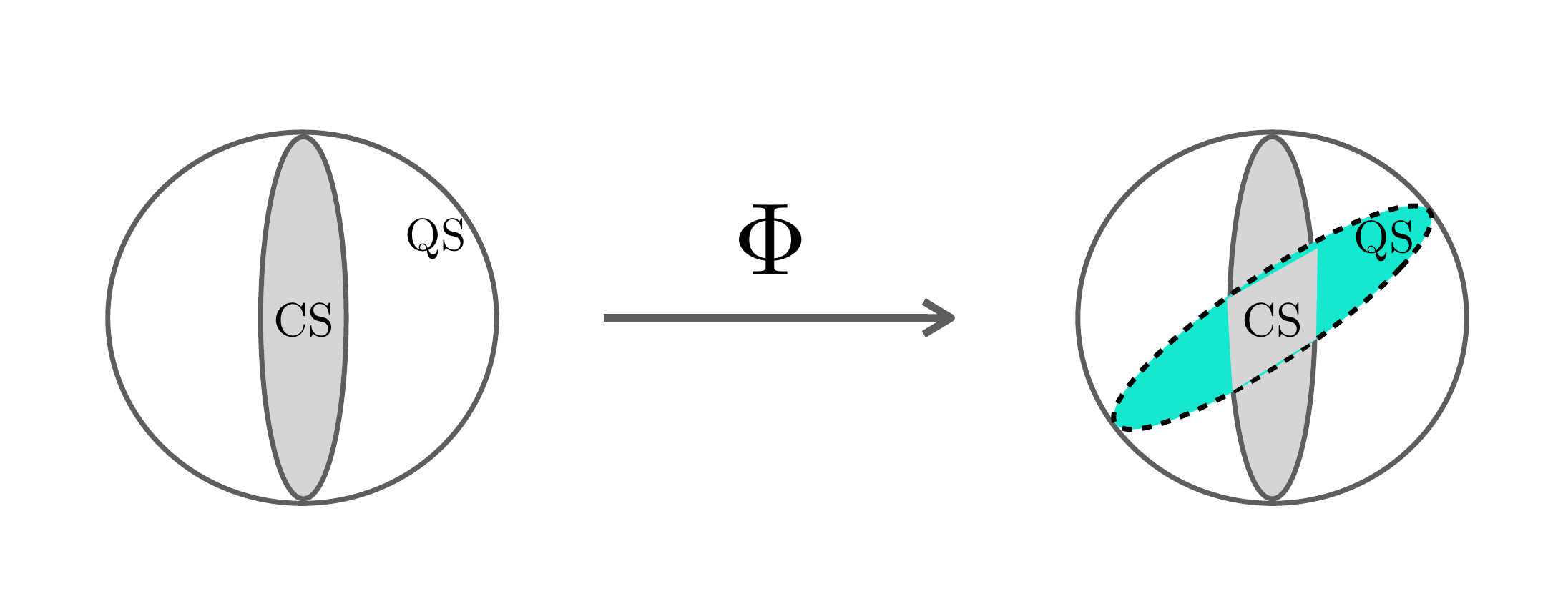}
\caption{Representation of the actuation of a quantum channel $\Phi$ on a given set of classical states (CS), defined as the eigenstates of a given observable $Q$, as well as their convex combinations. The set of quantum states (QS) is represented by the circle, which contains the classical set which respect to $Q$.  As a result of the invasiveness of $\Phi$, the set of classical states may be taken into a set containing nonclassical states,  in the area delimited by the shape with dashed border, in the right-hand side.  }\label{invasiveness}
\end{figure}

In Fig. \ref{invasiveness}, we represent the set of classical states (CS), given by the eigenstates of $Q$ and their convex combinations, as a subset of the set of quantum states (QS). 
An illustrative example of an invasive operation is given, and we see that some classical state are mapped into some nonclassical ones.

In resource theory, any monotone can be used as a quantifier for the specific resource under consideration.
In our case, we want to develop a resource theory for transformations.
In order to a function $I$ to be a quantifier of the invasiveness of $\Phi$ with respect to an observable $Q$, it is natural to demand

\begin{enumerate}
\item Positivity: $ I(\Phi) \ge 0 $ while $ I(\Phi_\text{Free})=0 $ for all $\Phi_\text{Free}$.
\item Monotonicity under free operations:
\begin{equation}
I(\Phi_{\text{Free}}\circ \Phi \circ \tilde{\Phi}_{\text{Free}}) \le I(\Phi),
\end{equation}
for all $\tilde{\Phi}_{\text{Free}}$ and $\Phi_{\text{Free}}$.
\item Convexity: $ I(\sum_i p_k  \Phi _k) \le \sum p_k I(\Phi _k)  $.
\end{enumerate}

For a fixed observable $Q$, one can define different classes of quantifiers, using different choices of classical states.
One possibility is to fix some classical state, $\gamma$, and to have (for each choice of $\gamma \in \Gamma$) a quantifier of the invasiveness of $\Phi$ as detected by $Q$ and $\gamma$. 
Another possibility is to optimize over all $\gamma \in \Gamma$.

Next, we present an example of a quantifier of invasiveness.
We will seek to make it meaningful in the context of quantum metrology by introducing a parameter $\theta$ in the scenario described above.
By doing so, the resource theory of invasiveness of operations can be connected to an application in the context of a quantum information protocol.
More generally, by focusing in finding meaningful quantifiers in the context of a given task, alternative routes to exploit and understand nonclassicality can be explored.
We believe that this could be an interesting approach to be taken into consideration in future investigations.

\section{Invasiveness as a resource to parameter estimation}\label{QuantInvasiveness}

The quality of the estimation of a given parameter $\theta$ of a physical system can be assessed through the evaluation of the Fisher information of the process. 
Specifically, the Fisher information $F$ determines the sensitivity of the estimation of a given parameter $\theta$, as it bounds the standard deviation $ \Delta \theta $ as $\Delta \theta \ge 1/\sqrt{\nu F(\theta)} $, where $\nu$ is the number of realizations of the experiment, assuming unbiased measurements. 
The Fisher information can be expressed as follows
\begin{equation}\label{eq2}
F(\theta)=\sum_l P_l(\theta)\left[\frac{\partial \ln P_l(\theta)}{\partial \theta}\right]^2,
\end{equation}
where $P_l(\theta)$ are the probabilities associated with each possible result $l$ of the measurement, satisfying $\sum_l P_l(\theta)=1$.

The generalization of the Fisher information to quantum mechanics can be done by writing $ P_l(\theta) = {\rm Tr}[\rho(\theta)E_l] $, where $\rho(\theta)$ depending on $\theta$ and $\{E_l\}$ is a \textit{positive operator valued measure}. 
The maximum value $\mathcal{F}_Q$ that $F(\theta)$ can assume, the {quantum Fisher information}  \citep{Helstrom, Holevo, Braunstein, Braunstein2}, can be obtained through the maximization over all quantum measurements,  $\mathcal{F}_Q = \max_{\{E_l\}}F(\rho,\{E_l\})$. 
As a result, the quantum Fisher information corresponds to the Fisher information associated with the optimal measurement, i.e. the one which gives the most precise estimation for $\theta$.

\begin{figure}[h]
\centering 
\includegraphics[width=8.5cm]{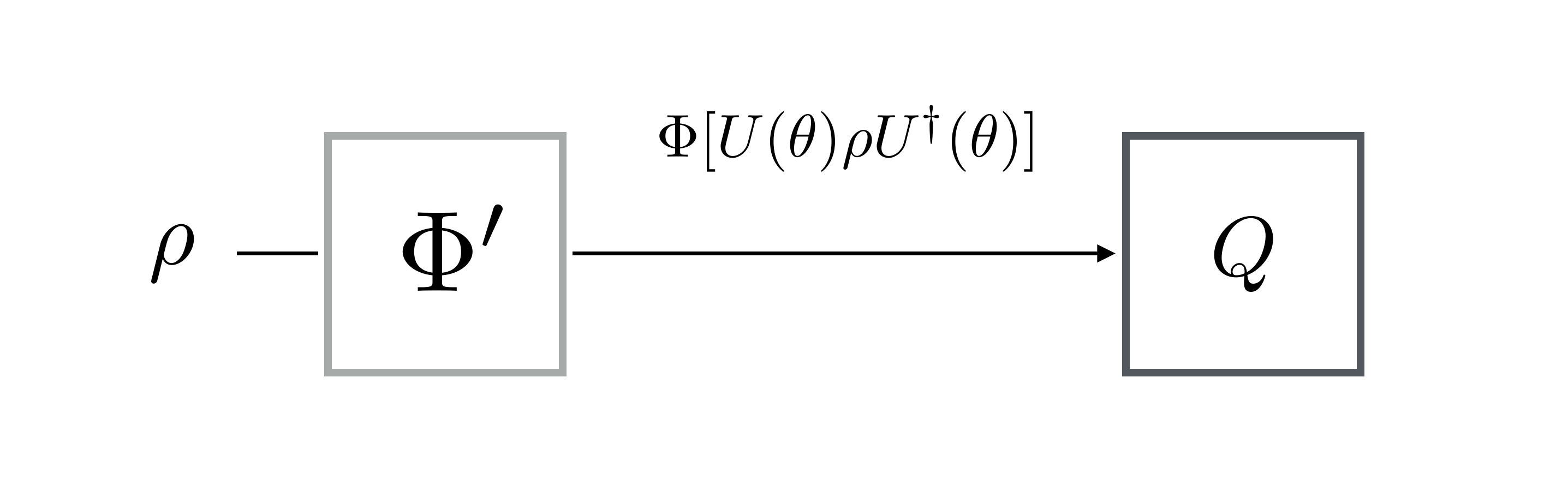}
\caption{The state $\rho$ is subjected to $\Phi^\prime_\theta = \Phi \circ \eta_\theta$: $\Phi$ the unitary transformation $\eta_\theta$ and subsequently $\Phi$ are applied to the state $\rho$, and the observable $Q$ is finally measured.  }\label{Scenario2}
\end{figure}

In relation with the resource theory of invasiveness defined above, we will show that the Fisher information is a suitable quantifier of the invasiveness of a class of $\theta$-dependent operation.
To do so, we consider the scenario described above, now with a unitary imprinting of a parameter $\theta$.
Consider a unitary transformation $U(\theta) = e^{-iA\theta}$, $A$ being a Hermitian operator, the role of which is to imprint the parameter $\theta$ on the system's state (see Fig.~\ref{Scenario2}). 
Let us denote $\eta_\theta$ the actuation of $U(\theta)$ on density operators: $\eta_{\theta} \rho = U(\theta)\,\rho\, U^{\dagger}(\theta)$.
Therefore, the transformation undergone by an arbitrary state $\rho$, which can or cannot be invasive, is now denoted by $\Phi^\prime_\theta=\Phi\circ\eta_\theta$.

In the classical scenario, we expect that $[A,Q] = 0$ is satisfied. As a result, an example of transformation $\Phi^\prime_\theta$ which cannot generate invasiveness is $\Phi_{\text{Free}}\circ[\eta_\theta]_{\text{Free}}$, where $[\eta_\theta]_{\text{Free}}$ corresponds to a unitary such that $[A,Q] = 0$. In particular, note that the power of the invasiveness of $\Phi$ becomes clear when we impose $[A,Q] = 0$.
In this case, without $\Phi$ or with some non-invasive $\Phi_{\text{Free}}$, the state $\rho_f (\theta)= \Phi_{\text{Free}} \circ [\eta_{\theta}]_{\text{Free}} \rho$ is also classical, and the probabilities are independent of $\theta$, giving null Fisher information for this process.

On the other hand, if $\Phi$ can transform a classical state $\rho^C$ into some non-classical one, \ie into some state which can not be written as a mixture of states with well defined value $q$, then interference fringes can show up in the probabilities as functions of $\theta$, giving rise to positive Fisher information.
 
By considering such a scenario and the definition of the quantifier of invasiveness $I$ above, we arrive at a result expressing the nonclassicality from the point of view of the Fisher information.
\\
\\
\begin{theorem}\label{theorem}
 \textit{A suitable quantifier of the invasiveness of a quantum channel $\Phi^\prime_\theta = \Phi \circ \eta_\theta$, where $\eta_\theta$ is a unitary transformation $U(\theta) = e^{-i\theta A}$, given  an arbitrary state $\rho$, $\rho_f = \Phi^\prime_\theta(\rho)$ and the POVM $\{\ket{q}\bra{q}\}$, is}
\begin{equation}\label{Quantifier}
I(\Phi^\prime_\theta) = F_{\rho_f}(\theta),
\end{equation}
\textit{where $F_{\rho_f}(\theta)$ is the Fisher information of the state $\rho_f$}.
\end{theorem}
\
\begin{proof}
In order to prove condition 1, note that a free operation can only permute or coarse-grain the diagonal elements of $\rho$.
Moreover, by assumption, the only dependence that $\Phi^\prime_\theta = \Phi \circ \eta_\theta$ can have on the parameter $\theta$ comes from $\eta_\theta$.
Consistently, any free operation with a $\theta$ dependence can be factored as  $\Phi_{\text{Free}}\circ[\eta_\theta]_{\text{Free}}$, with a $[\eta_\theta]_{\text{Free}}$ generated by some $A$ such that $[A,Q]=0$.
 As a result, the diagonal elements of $\rho_f$ will not depend on $\theta$.
 Therefore, $I(\Phi^\prime_\text{Free}) \equiv F_{\rho_f}(\theta) = 0$.
\\
Condition 2 is also satisfied: the relevant terms are on the diagonal of $\rho_f$, since we are interested in the probabilities $P_{\ket{q}}(\theta) = {\rm Tr}(\ket{q}\bra{q}\Phi^\prime_\theta(\rho_f))$ associated with the eigenstates $\ket{q}$ of $Q$. 
Since free operations $\Phi_{\text{Free}}$ are generated by permutations of $\ket{q}$, Birkhoff theorem implies $\Phi_{\text{Free}}\left(\rho\right)$ is a convex combination of permutations of $\rho$, and convexity (condition 3) shows that it cannot increase $I\left(\Phi'\right)$.
Analogously, when calculating $I\left(\Phi_{\text{Free}} \circ \Phi'\right)$, one can consider the adjoint channel $\Phi_{\text{Free}}^{\dagger}$ acting on $\ket{q}\bra{q}$ to conclude that the new $P_{\ket{q}}\left( \theta\right)$ will be convex combinations of the old ones, with convexity closing the argument again.  
\\
Finally, condition 3 is fulfilled since $F_{\rho_f}(\theta)$ is convex - see Appendix \ref{appendixA}.
 \end{proof}
 
As a simple example, consider a qubit system, the eigenstates of $\sigma_z$ are denoted by $\{\ket{0},\ket{1}\}$ and considered classical, i.e. $Q=\sigma_z$. If $A=\sigma_x$, $\Phi=\mathbb{1}$, and $\rho=\ket{\psi}\bra{\psi}$ with $\ket{\psi}=\ket{0}$ in Fig. \ref{Scenario2}, we obtain
\begin{equation}\label{psif}
\ket{\psi_f}=\Phi^\prime_\theta(\ket{\psi})=e^{-i\theta\sigma_x}\ket{0}=\cos\theta\ket{0}-i\sin\theta\ket{1}.
\end{equation}
The probabilities of obtaining each of the outcomes associated with the eigenstates of $Q$ are $P_0(\theta)=|\braket{0}{\psi_f}|^2=\cos^2\theta$ and $P_1(\theta)=|\braket{1}{\psi_f}|^2=\sin^2\theta$. As a result,
\begin{multline}
I(\Phi^\prime_\theta)=F_{\ket{\psi_f}}(\theta)= \frac{1}{P_0(\theta)}\left[\frac{\partial  P_0(\theta)}{\partial\theta}\right]^2 +  \frac{1}{P_1(\theta)}\left[\frac{\partial  P_1(\theta)}{\partial\theta}\right]^2 = \\
\left[\frac{-2\cos\theta\sin\theta}{\cos\theta}\right]^2 + \left[\frac{2\sin\theta\cos\theta}{\sin\theta}\right]^2=4.
\end{multline}
This is actually a particular example of the more general case where $A=\cos\alpha\sigma_x+\sin\alpha\sigma_z$. In this case, the calculation of $I(\Phi^\prime_\theta)$ gives
\begin{equation}\label{IGeral}
I(\Phi^\prime_\theta)=\frac{4\sin^2\theta\cos^2\alpha}{1-\cos^2\theta\cos^2\alpha}.
\end{equation}
As a result, the particular case above with $A=\sigma_x$ (for $\alpha=0$, for example) is the most invasive of this family. It is interesting to note that invasiveness decreases monotonically in the interval $0 \le \alpha \le \pi/2$, and only in the extremal cases $\alpha=0,\pi/2$, it is independent of $\theta$.

It is important to stress that, in Theorem \ref{theorem}, while $\Phi$ is an arbitrary quantum channel which does not depend on $\theta$, $\Phi^\prime_\theta$ is a specific class of quantum channels, which consider the unitary imprinting of the parameter $\theta$.
Other dependences on $\theta$ could be considered, including the extreme case where all the dependences come from $\theta$-dependent free transformation which are not related to invasiveness - pretty much on the contrary, they are in the scope of classical metrology.
An important question yet to be tackled is to determine the most general class of $\theta$-dependent channels where the Fisher information is related to invasiveness.

\section{Discussion}\label{discussion}

As mentioned above, the connection between the notion of measurement invasiveness and the violation of NSIT conditions and LGIs has been studied in Refs. \citep{Frowis,Moreira2}. 
Specifically, in Ref. \citep{Frowis}, states with large quantum Fisher information are associated with the violation of NSIT conditions for large measurement uncertainties. 
In turn, the connection between LGI violation and optimal scenarios in quantum metrology has been investigated in Ref. \citep{Moreira}. Here, by utilizing a generalized definition of quantum invasiveness, we showed that the Fisher information is a quantifier of quantum invasiveness for a certain class of quantum channels. 
This result allows us to establish a direct association between quantum invasiveness and sensitivity, as the first can be seen as a resource for the latter.

This paper also clarifies the relationship between invasiveness and coherence.
As pointed out in Refs. \citep{Baumgratz, Yadin2}, classes of coherence quantifiers of a state $\rho$ can be found from distance quantifiers $D$ fulfilling the following two properties, with $ \tau \in \mathcal{I}$ being incoherent states: $ D(\rho,\tau) = 0 $, if and only if $ \rho = \tau $ and $D$ is contractive under trace preserving quantum operations $\Lambda$, i.e. $ D(\Lambda(\rho),\Lambda(\tau)) \le D(\rho,\tau) $. 
Thus, the quantifier of coherence associated with $D$ is~$ C_D = \min_{\tau \in \mathcal{I}} D(\rho_C,\tau) $. As remarked above, a distance between $\rho$ and $\rho_f$ different from zero is necessary for invasiveness. 
In this way, in the scenario of Fig. \ref{Scenario}, whenever the input state is an incoherent state $ \tau \in \mathcal{I} $, invasiveness means that coherence, as defined within the framework of resource theories, is generated with respect to the basis of $Q$.

Moreover, Eq. \eqref{Quantifier} provides us with a insightful fashion of quantifying nonclassicality in Leggett-Garg scenarios, as a nonzero value for the invasiveness quantifier $I(\Phi)$ will be necessarily due to the distinction between quantum and classical scenarios from the point of view of the Fisher information. 

In summary, based on a generalized definition of invasiveness of quantum operations, we have proposed a resource theory of quantum invasiveness. 
Within the resource theoretic framework, free states are considered to be classical states and therefore associated with the eigenstates and their convex combinations of the measurement observable. 
In turn, free operations are those mapping classical states into classical states.
In this context, we showed that the Fisher information is a quantifier of quantum invasiveness of a class of quantum channels. This result sheds light on the utility of quantum invasiveness within the context of protocols in quantum metrology as quantum invasiveness can be considered a resource for sensitivity. 
We have also seen that the proposition of a resource theory of invasiveness allows us to establish a connection to the framework of the resource theory of coherence \citep{Baumgratz}. 
In this perspective, we expect that the proposed resource theory of quantum invasiveness may lead to new insights and improvements concerning the implementation of metrological protocols.
 Among the questions evoked by this work - which are certainly worth further investigation - are the determination of the tasks for which invasiveness can be useful, as well as the question of the interconvertibility between invasiveness and other nonclassical resources such as coherence and purity \citep{Chuang}, given that resources do not necessarily compete.

\section*{Acknowlegments}

The authors acknowledge the anonymous Referees for useful comments. S. V. M. acknowledges financial support from the Brazilian agency CAPES. M. T. C. acknowledges partial support from the Brazilian agency CNPq. This work is part of the Brazilian National Institute of Science and Technology on Quantum Information.

\appendix

\section{}\label{appendixA}

Here, by following the steps of Ref. \citep{Cohen}, we provide a demonstration that the Fisher information satisfies condition 3. Therefore, we must be able to prove that
\begin{equation}\label{convexity2}
F((1-\gamma)\rho_1(\theta) + \gamma \rho_2(\theta)) \le (1-\gamma) F(\rho_1(\theta)) + \gamma F(\rho_2(\theta)),
\end{equation}
where $\rho_1(\theta)$ and $\rho_2(\theta)$ are two arbitrary quantum states depending on the parameter $\theta$ and $ 0 \le \gamma \le 1 $. 

Before proceeding, it will be useful to rewrite \eqref{eq2} as
\begin{equation}\label{Fisher2}
F(\theta)=\sum_l \frac{1}{P_l(\theta)}\left[\frac{\partial P_l(\theta)}{\partial \theta}\right]^2.
\end{equation}

Let $ P_l(\theta) $ and $ G_l(\theta) $ be the probabilities associated with a given outcome $l$ when measuring $ Q(\theta) $ for states $\rho_1(\theta)$ and $\rho_2(\theta)$, respectively. Assume that for a given $l$, one can write the following inequality
\begin{equation}\label{inverse}
\frac{[ (1-\gamma)P^\prime_l(\theta) + \gamma G^\prime_l(\theta) ]}{(1-\gamma)P_l(\theta) + \gamma G_l(\theta)}^2 > (1-\gamma) \frac{P^\prime_l(\theta)^2}{P_l(\theta)} + \gamma \frac{G^\prime_l(\theta)}{G_l(\theta)}^2,
\end{equation}
where $ P^\prime_l(\theta) $ and $ G^\prime_l(\theta) $ are the derivatives  of $P_l(\theta)$ and $G_l(\theta)$, respectively.
By simplifying \eqref{inverse}, we obtain
\begin{multline}
2\gamma(1-\gamma)P^\prime_l(\theta) G^\prime_l(\theta) P_l(\theta) G_l(\theta) > \\ \gamma(1-\gamma)[P^\prime_l(\theta)^2 G^2_l(\theta) + G_l^\prime(\theta)^2 P_l^2(\theta)],
\end{multline}
which finally gives
\begin{equation}
0 > (P_l (\theta) G_l^\prime (\theta) - G_l (\theta) P_l^\prime (\theta) )^2,
\end{equation}
which cannot be satisfied for any $l$. Thus, we can conclude that for all $l$, the following inequality is true
\begin{equation}\label{convexity}
\frac{[ (1-\gamma)P^\prime_l(\theta) + \gamma G^\prime_l(\theta) ]}{(1-\gamma)P_l(\theta) + \gamma G_l(\theta)}^2 \le (1-\gamma) \frac{P^\prime_l(\theta)^2}{P_l(\theta)} + \gamma \frac{G^\prime_l(\theta)}{G_l(\theta)}^2.
\end{equation}

By summing up both sides of \eqref{convexity} over $l$ and by referring to \eqref{Fisher2}, we finally obtain \eqref{convexity2}.

\end{document}